\documentclass[11pt]{article} 
\usepackage{cospar} 
\usepackage{url} 
 
\usepackage{graphicx}
 
\newcommand{\la}{\mbox{ \raisebox{-.5ex}{$\stackrel{\textstyle <}{\sim}$} }} 
\newcommand{\ga}{\mbox{ \raisebox{-.5ex}{$\stackrel{\textstyle >}{\sim}$} }}

\title{VISIBILITY-BASED DEMODULATION OF RHESSI LIGHT CURVES} 
 
\author{K. Arzner\address{Paul Scherrer Institut,  
    CH-5232 Villigen PSI, Switzerland}}  
 
\begin{document} 
 
\maketitle 
 
\begin{abstract} 
The Reuven Ramaty High Energy Spectroscopic Solar Imager (RHESSI) uses the 
rotational modulation principle (Schnopper et al., 1968) to observe temporally, 
spatially, and spectrally resolved hard X ray and gamma ray images of solar flares. 
In order to track the flare evolution on time scales that are commensurate with modulation, the 
observed count rates must be demodulated at the expense of spatial information. 
The present paper describes improvements of an earlier demodulation 
algorithm, which decomposes the observed light curves into intrinsic source  
variability and instrumental modulation.
\end{abstract} 
 
 
\section*{INTRODUCTION} 
 
The RHESSI (Lin et al. 2003) instrument is designed to explore particle acceleration in solar
flares (e.g., Miller et al., 1997). It observes photons at energies from 3 keV to 17 MeV
with spectral resolution up to $E/\Delta E$$\sim$500. In order to obtain images, RHESSI
is equipped with 9 pairs of X ray shadowing grids 
(9 `subcollimators'), which are fixed on the rotating spacecraft (spin period $T_S$$\sim$4s). 
The instantaneous transmission probability of one subcollimator, projected onto  
the solar disc (heliocentric cartesian coordinates ${\bf x}$), is called the `modulation pattern' 
(Hurford et al., 2003a). It has the approximate form ($i$=1...9) 
\begin{equation} 
M^i({\bf x},t) = a_0^i + a_1^i \cos \Big( {\bf k}^i(t) \cdot ({\bf x} - {\bf P}(t)) 
+ \psi^i \Big). 
\label{modpat} 
\end{equation} 
The wave vectors ${\bf k}^i(t)$ correspond to periods (`angular pitches') of  
$\sim$$2.6 \cdot 3^{i/2}$ asec, and ${\bf P}(t)$ is the imaging (optical) 
axis. The angle $\Phi^i_{\rm k}(t)$ between ${\bf k}^i(t)$ and the heliocentric $x$-axis 
is given by $\Phi^i_{\rm k}(t) = \pi/2 - \Phi_{\rm roll}(t) - \Phi^i_{\rm grid}$, 
where $\Phi_{\rm roll}(t)$ is the roll angle and
$\Phi^i_{\rm grid}$ is the grid orientation on the spacecraft. The 
triples $(\Phi_{\rm roll}(t),{\bf P}(t))$ are referred to as `aspect solution', and  
are continuously monitored by the on-board aspect systems (Fivian and Zehnder, 2003;  
Hurford et al., 2003b). The coefficients $a_0^i,a_1^i,\psi^i$ in eq. (\ref{modpat}) depend on energy 
and (weakly) on ${\bf x}-{\bf P}(t)$; they describe the X ray transmission of the grids. 
Above 100 keV the grids become increasingly transparent. In what follows, the energy band  
is fixed, and the energy dependence is omitted. 
 
At any time, the 9 subcollimators respond to 9 different Fourier components of the  
solar brightness distribution. During half a RHESSI rotation, 
up to thousand Fourier components are collected, from which 
the spatial brightness distribution can be restored (Hurford et al., 2003a). The imaging 
information is thus encoded in the temporal modulation of the observed count rates.  
In order to track impulsive events (energy release, acceleration) on  
time scales below 2s, the observed count rates must be demodulated at the 
expense of spatial information. 

The present article reports on further development of the demodulation method of Arzner (2003), 
which compared count rates with similar modulation patterns. This constraint was rather restrictive,
since both the roll angle and the grid phase had to agree. The request for grid
phase agreement can in fact be relaxed by the use of visibilities\footnote{At each time, the two
`visibilities' are the projections of the instantaneous brightness distribution 
onto sine- and cosine components of the modulation pattern.}; 
this is described in Section 2 below. In order to work properly, this approach requires that
the time binning is synchronized with the spacecraft rotation, i.e., that $\Phi_{\rm roll}$ 
(mod $\pi$) recurs after an integer number of time bins $\Delta t$. 
On the other hand, $\Delta t$ must be an integer
multiple of a binary micro second, which is the hardware time resolution. The improved
version choses an optimum compromise between the two conflicting constraints, while
keeping the number of counts per bin at a statistically uselful level ($\sim$ 10).
Section 3 shows benchmark simulations and the demodulation of real RHESSI data.

\section*{DEMODULATION METHOD} 
 
The goal of the method is to separate spatial and temporal variations of the solar brightness  
distribution. The basic observation is that the visibilities at equal grid 
orientation should be compatible if the brightness distribution did not change with time.
Deviations are attributed to time dependence of the source distribution. Technically,
this is done by a regularized maximum-likelihood fit with many ($>$10$^2$) parameters describing
both spatial and temporal source variability.

\begin{figure} 
\centerline{\includegraphics[width=150mm]{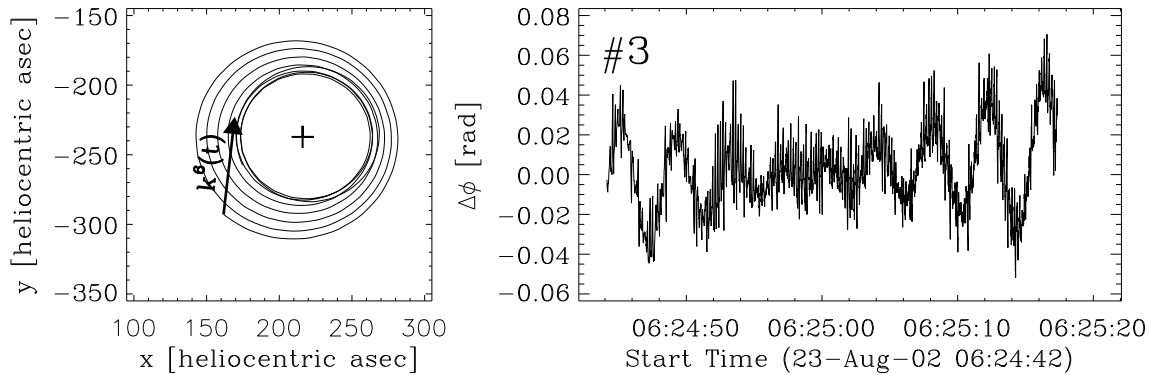}}
\caption{Left: Motion of the imaging axis ${\bf P}(t)$ (circles) during $N_P$ = 8 spacecraft revolutions 
at the Flare of August 23, 2002, 06:25 UT. 
The wave vectors (arrow indicates orientation of ${\bf k}^6$ at final time)  
rigidly co-rotate with ${\bf P}(t)$. The average spin axis is marked by a cross. 
Right: change of modulation phase of subcollimator 3 per $\Delta t$ at the average spin axis
($\Delta \phi = \dot{\phi} \, \Delta t$). The condition
$|\Delta \phi | \ll 2 \pi$ is the basis of Eq. (\protect\ref{appr}).} 
\label{aspect} 
\end{figure} 

Let us start with a brief description of the spacecraft motion, which determines the time-dependent 
instrumental response (Eq. \ref{modpat}). During a few ($N_P$$\la$10) spin periods, the RHESSI motion is 
composed of an approximately uniform rotation and a slow translation (Figure \ref{aspect} left). 
We formalize this by making the following assumptions (which are checked from the
aspect solution): (i) the spin period $T_S$ is constant to accuracy $\Delta t$ over the
time of interest ($N_P T_S$); (ii) the wave vectors ${\bf k}^i(t)$ rotate
clockwise with spin period $T_S$; (iii) the imaging axis ${\bf P}(t)$, which is generally not aligned 
with the spin axis\footnote{Strictly speaking, ${\bf s}(t)$ denotes the
penetration point of the instantaneous RHESSI spin axis through the solar disk.} ${\bf s}(t)$, rotates (clockwise, period $T_S$) 
around ${\bf s}(t)$. (iv) the spin axis and  
$|{\bf s}(t)$--${\bf P}(t)|$ vary slowly due to precession ($f_{prec} \sim$ 0.015 Hz), inertia  
adjustment, and magnetic torquing (Lin et al., 2003).
  
We next set up a model for the solar brightness distribution. It is assumed that the
true brightness can be represented as
\begin{equation} 
B({\bf x},t) = \sum_k r_k(t) B_k({\bf x}) 
\label{model} 
\end{equation} 
where $B_k({\bf x})$ are arbitrary non-negative normalized ($\int \hspace{-.3mm} B_k ({\bf x}) d {\bf x}$=1)
functions and $r_k(t)$ (ct/s) is the instantaneous count rate.
In principle, Eq. (\ref{model}) describes all possible brightness distributions,
but in practice we restrict ourselves to a few source components $k$; perhaps, a gradual plus an
impulsive component with different intrinsic time scales. Impulsive components could be
associated with magnetic footpoints. It should be pointed out that Eq. (\ref{model}) 
cannot account for continuously moving sources, and that non-solar background
is neglected. 

Combining the RHESSI response (Eq. \ref{modpat}) with the brightness model (Eq. \ref{model})
one can predict the expected counts in time bins $\Delta t$. 
As we wish to resolve the source evolution it is assumed that $r_k(t)$ 
varies slowly during $\Delta t$. The expected number of counts in the $i$-th subcollimator
and $t$-th time bin is then given by
\begin{equation} 
\lambda^i_t = L^i_t \sum_{k=0}^1 r_{k,t} m^i_{k,t} 
\label{lambda} 
\end{equation} 
where the livetime measure $L^i_t$ accounts for detector dead time 
(Smith et al., 2003) and data gaps\footnote{Data gaps occur independently 
in different subcollimators and have durations of milliseconds to seconds 
(Figure \protect\ref{demod_result}c). They affect some 30\% of all data, and
are presumably caused by cosmic rays.
The livetime measures $L^i_t$ are between zero and one.}, $r_{k,t}$ = $(\Delta t)^{-1} \int_t^{t+\Delta t} 
dt' \, r_k(t')$ is the (discrete) count rate, and
\begin{equation}
m^i_{k,t} = \int_t^{t+\Delta t} \hspace{-3mm} dt' \int  \hspace{-1mm} d{\bf x} \, 
M^i({\bf x},t') B_k({\bf x}) \, . \label{x} \\ 
\end{equation}	
We recall that ${\bf x}$ denotes heliocentric coordinates. 
While being natural, these coordinates
are not well adapted to the observing geometry, and do not allow to separate fast (rotational)
and slow (translational) contributions to modulation. Such a separation is, however, possible
if we set ${\bf x}$ = ${\bf x}'+ \langle {\bf s} \rangle$, where $\langle {\bf s} \rangle$ is the average
spin axis (Figure \ref{aspect} left cross; average over $N_P$ spin periods). After expanding the cosine 
in Eq. (\ref{modpat}), Eq. (\ref{x}) becomes
\begin{eqnarray} 
m^i_{k,t}& = & \int_t^{t+\Delta t} \hspace{-3mm} dt' \int \hspace{-1mm} d{\bf x}'  
		\bigg( a_0^i(t') + a_1^i(t') \cos ( {\bf k}^i(t') \cdot {\bf x}')  
		\cos \Big\{ {\bf k}^i(t') \cdot (\langle{\bf s}\rangle - {\bf P}(t')) + \psi^i(t')) \Big\} \nonumber \\	 
	&   &	- a_1^i(t') \sin ( {\bf k}^i(t') \cdot {\bf x}')  
		\sin \Big\{ {\bf k}^i(t') \cdot (\langle{\bf s}\rangle - {\bf P}(t')) + \psi^i(t') \Big\} \bigg)
		B_k({\bf x}'+\langle{\bf s}\rangle) \, .\label{x'}
\end{eqnarray}	
Now, the term in curly brackets is the grid phase at the average spin axis, which is a slowly varying quantity. 
If also $a^i_0(t)$ and $a^i_1(t)$ are weakly varying, then Eq. (\ref{x'}) can be approximated by
\begin{eqnarray}	
m^i_{k,t}& \simeq & a_{0,t}^i \Delta t + a_{1,t}^i \cos \phi^i_t 
		\int_t^{t+\Delta t} \hspace{-3mm} dt' \int d{\bf x}' \cos ( {\bf k}^i(t') \cdot {\bf x}') B_k({\bf x}'+\langle{\bf s}\rangle) \nonumber\\		  
	&   &	- a_{1,t}^i \sin \phi^i_t 
		\int_t^{t+\Delta t} \hspace{-3mm} dt' \int d{\bf x}' \sin ( {\bf k}^i(t') \cdot {\bf x}') B_k({\bf x}'+\langle{\bf s}\rangle) \label{appr} \\ 
	& \doteq & a_{0,t}^i \Delta t + a_{1,t}^i \cos \phi^i_t \, C^i_{k,t} - a_{1,t}^i \sin
		\phi^i_t \, S^i_{k,t} \label{CS}		 
\end{eqnarray} 
where $a_{n,t}^i$ is a discrete version of $a_n^i(t)$, $\phi^i_t = {\bf k}^i_t \cdot (\langle{\bf s}\rangle-{\bf P}_t) + \psi^i_t$
is the discrete grid phase at average spin axis,
and the last line defines the ($\Delta t$-integrated) visibilities $C^i_{k,t}$ and $S^i_{k,t}$.
The visibilities satisfy $(C^i_{k,j})^2 + (S^i_{k,j})^2$ $\le$ $(\Delta t)^2$, $C^i_{k,j+N_S/2}$ = $C^i_{k,j}$, and  
$S^i_{k,j+N_S/2}$ = $-S^i_{k,j}$, where $N_S$ is the (even) number of time bins per spin period.
The visibility approach (Eqns. \ref{lambda} and \ref{appr}) is valid if 
$|\dot{r}_k/r_k| \Delta t \ll 1$ and
\begin{equation}
|\dot{a}_0^i/a_0^i| \Delta t \ll 1, \;\;\;
|\dot{a}_1^i/a_1^i| \Delta t \ll 1, \;\;\;
(2 \pi)^{-1} |\dot{\phi}^i| \Delta t \ll 1 \, ,
\label{cond}
\end{equation}
which must hold during the whole time interval $N_P T_S$.
A safe threshold for Eqns. (\ref{cond}) is 0.1, which
is checked by the computer code. This is usually uncritical (Figure \ref{aspect} right): from typical aspect data, one finds 
that time bins as large as $\Delta t$ = 0.3s are admissible even for the finest subcollimator.
It should be stressed that this is much larger than the modulation period 
-- the modulation must not be resolved to 
predict similar ($\Delta t$-integrated) visibilities. Of course, integrating over modulation razes 
spatial information, but this does not affect the time profiles $r_{k,t}$.

At this point we have completed the forward model (Eqns. \ref{lambda},\ref{CS}).
It has parameters $\{C^i_{k,j}, S^i_{k,j},r_{k,t}\}$, where
1 $\le$ $j$ $\le$ $N_S/2$, 1 $\le$ $t$ $\le$ $N_S \hspace{-.6mm} \cdot \hspace{-.6mm} N_P$, 
and $i$ is out of a subcollimator set satisfying Eq. (\ref{cond}).
We turn now to the counting statistics. Let the observed light curve of subcolllimator $i$
have $c^i_t$ counts in time bin $t$. These counts should scatter around the expectation value
$\lambda^i_t$ (Eq. \ref{lambda}) according to Poisson statistics. The agreement between observation and 
prediction is therefore measured by the Poisson log likelihood ratio (Eadie et al., 1971) (`C-statistic')
\begin{equation} 
{\rm log \, {\sf L}} = {\sum_{it}}^* \Big\{ - \lambda^i_t + c^i_t 
\Big(1 + \ln \frac{\lambda^i_t}{c^i_t} \Big) \Big\}  \, .
\label{logL}  
\end{equation} 
The likelihood ${\sf L}$ is proportional to the probability that $\{ c^i_t \}$ is observed if $\{ \lambda^i_t \}$
was true. Log {\sf L} is negative, and reduces to $-\frac{1}{2} \chi^2$
in the limit of high count rates.
The asterisks in Eq. (\ref{logL}) indicates that the sum is restricted to count rates with livetimes  
$L^i_t$ $\ge$ $L_{\rm min}$, with $L_{\rm min}$ typically chosen as 0.5. This prevents from 
pile up and from redundant numerical operations, since 0 $<$ $L^i_t$ $<$ $L_{\rm min}$ 
is rare but $L^i_t$ = 0 is frequent ($\ga 25$\% data gaps).  
The rejection of low livetimes is not expected to affect the result. 

Since there are more model parameters $\{C^i_{k,j}, S^i_{k,j},r_{k,t}\}$ than observations $\{c^i_t\}$,
and since the model parameters are not independent of each other, the maximum-likelihood 
problem does not have a unique solution. In order to regularize the problem we assign an a priori probability 
$\exp \{ - \frac{\alpha_k}{2} \sum_t (r_{k,t+1}-r_{k,t})^2 \}$ to the $k$-th time profile $r_{k,t}$.
This choice favors smooth time profiles, and allows an explicit trade-off between the smoothness 
of the demodulation and the likelihood. It also has the useful feature that it 
linearly interpolates across data gaps. Taking the a priori probability
into account, the logarithm of the total (observational + a priori) probability ${\sf P}$ 
becomes, according to Bayes' rule,
\begin{equation} 
{\rm log \, {\sf P}} = {\rm log \, {\sf L}} - \frac{1}{2} \sum_{kt} \alpha_k (r_{k,t}-r_{k,t-1})^2 \, . 
\label{F}  
\end{equation} 
The quantity $\log \, {\sf P}$ possesses a unique maximum which is found by Newton/Marquardt type iterations. 
The coefficients $\alpha_k$ determine the smoothness of the solutions $r_{t,k}$.
Small $\alpha_k$ yield small correlation times $\tau_k$ of $r_{t,k}$. While the
exact relation between $\alpha_k$ and $\tau_k$ depends on the actual data set
and on the definition of $\tau_k$, there exists a useful empirical estimate,
\begin{equation}
\tau_k \sim \max (\Delta t, \sqrt{\alpha_k} \, \langle c \rangle / \langle a_0 L \rangle  ) \, .
\label{tau} 
\end{equation}
In Eq. (\ref{tau}), $\tau_k$ is defined via the normalized
autocorrelation $A_k(t)$ of $r_{t,k}$ by $A_k(t) \simeq 1 - \frac{1}{2} (t/\tau_k)^2 + O(t^4)$.
The accuracy of Eq. (\ref{tau}) is within a factor two. The adjustment of $\alpha_k$ is
done by the user. The Newton/Marquardt iterations have the property that two solutions $r_{t,k}$ 
agree if their $\alpha_k$ agree.

\begin{figure}[ht] 
\includegraphics[width=95mm]{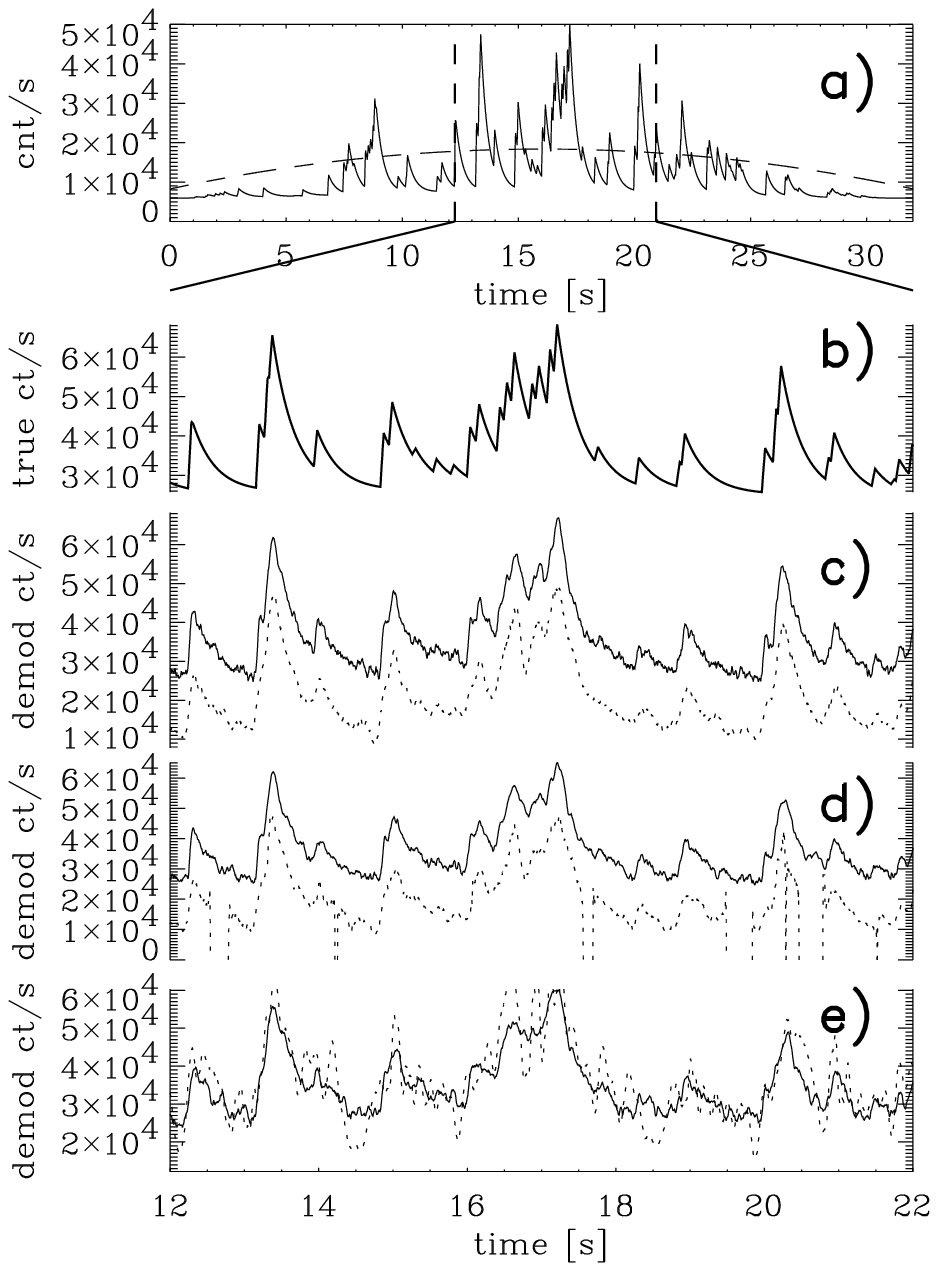}
\begin{minipage}{85mm}
\vspace{-11.5cm}
\includegraphics[width=85mm]{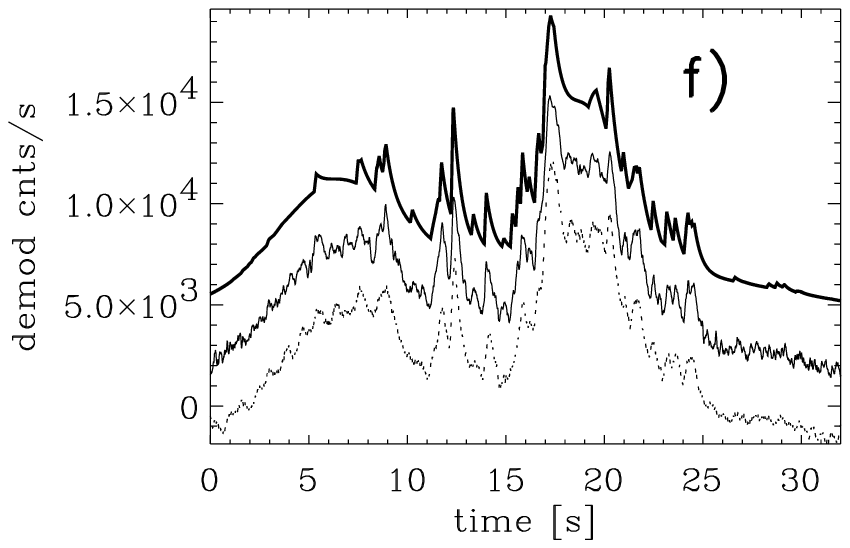} \\[8mm]
\includegraphics[width=85mm]{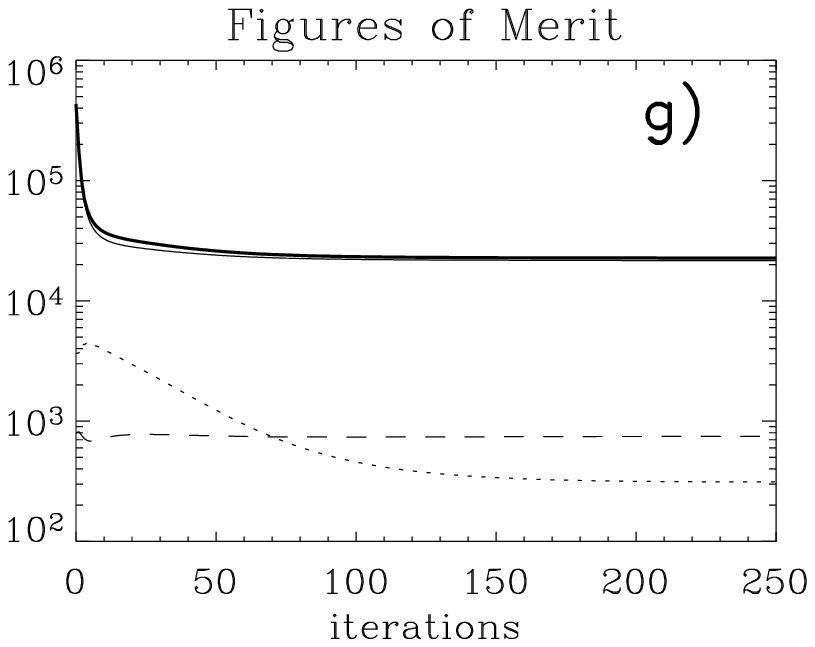}
\end{minipage}
\vspace{-8mm}
\caption{Benchmark simulations: a) true time profiles of impulsive (solid) and gradual (dashed) sources; b) 
close-up; c) demodulation from subcollimators 1-9 (solid), compared 
to an average over subcollimators 1-3 and time $\tau_a$=0.1s (dotted, shifted for better clarity);  
d) similar to c) including 30\% simulated data gaps; e) demodulation 
(solid) and average (dotted, $\tau_a$=0.1s) of subcollimators 7-9 (no data gaps). 
f): robustness against violation of Eq. (\protect\ref{model}) -
top to bottom curves: true unmodulated light curve of 5 independent components; two-component demodulation from 
subcollimators 1-9; average over subcollimators 
1-3 and time $\tau_a$=0.25s. The middle and bottom curves are shifted for better clarity.
g): iterative convergence of Eq. (\protect\ref{F}): $-\log {\sf P}$ (boldface); $-\log {\sf L}$ (solid);
gradual ($k$=0, dotted) and impulsive ($k$=1, dashed) smoothness constraints.} 
\label{sim} 
\end{figure}

\section*{RESULTS AND DISCUSSION} 
 
As a benchmark, Figure \ref{sim}a-e shows simulations where the model assumption (Eq. \ref{model}) 
is met. There are two sources at (-715'',630'') and (-650'',690''), of size 1'' and 3'', with impulsive  
(Figure \ref{sim}a solid line) and gradual (Figure \ref{sim}a dashed) time profiles. 
The small source sizes yield maximal modulation and therefore represent a worst-case
test for demodulation and residual modulation.
Fig \ref{sim}b shows a close-up of the true impulsive time profile, which is to be 
compared with the estimates c)-e) obtained from simulated `observations' with 
$\langle c^i_t \rangle/\Delta t$ = $6 \cdot 10^{3}$. The simulated aspect solution has the form
${\bf P}(t)$ = ${\bf v}_{\rm prec} t + r_{\rm cone} (\cos (\delta-\Omega t), \sin(\delta-\Omega t))$
with ${\bf v}_{\rm prec}$ = (0.8,1.3) asec/s the instantaneous precession rate,
$\Omega$ = $2 \pi / 4s$ the RHESSI angular frequency, $r_{\rm cone}$ = 120'' the coning
radius, and $\delta$ a phase offset. Figure \ref{sim}c-d shows the demodulation 
results (solid line) and averages over subcollimators 1-3 and time (dotted line) with 
similar nominal resolution.
In Figure \ref{sim}c, $\Delta t$=5ms, and the demodulation $r_{0,t}$+$r_{1,t}$ (solid line) of 
subcollimators 1-9 is compared with a moving average $\langle c \rangle/\langle a_0 L \rangle$  
(dotted) over subcollimators 1-2 and time interval $\tau_a$=0.1s. The autocorrelation time 
constant of the impulsive component $r_{1,t}$ is $\tau_1$ = 0.09s, while Eq.
(\ref{tau}) gives $\tau_1$ $\sim$ 0.14s. Numerically, the rms deviations from the true time profile are 5\% (average) and 3\% 
(demodulation), while the statistical error assigned to the $\Delta t$-binned true light curve
would be about 6\%. While the benefit from 3\% to 5\% is rather small and within the
statistical error, the situation changes if data  
gaps are present. Figure \ref{sim}d is similar to \ref{sim}c, but includes simulated data gaps 
with $\langle L^i_t \rangle$=0.7. Simultaneous data gaps in subcollimators 1-3 
with duration $>$$\tau_a$ cannot be interpolated with a moving average (dotted line), 
and the rms deviation of the moving average from the true time profile is 25\% 
(whereas it is 4\% for the demodulation). If one tries to improve the moving average
by including coarser subcollimators one encounters the difficulty that
their (at least, slowest) modulation commensurates with the time scale to be resolved. The moving
average then fails even in the absence of data gaps (Figure \ref{sim}e dotted). 
A possible way out is demodulation as described in this paper (Figure \ref{sim}e solid). 
Here, both the demodulation (solid) and average (dotted) involve subcollimators 7-9.
The rms deviation from the true time profile is 6\% (demodulation) and 18\% (average),
respectively.
A test was also made for violation of the model assumption (\ref{model}).
Figure \ref{sim}f shows the demodulation of (simulated) 
counts from 5 sources with different time profiles. The demodulation (Figure \ref{sim}f middle curve) uses subcollimators 
1-9 and is based on the (wrong) assumption that $B({\bf x},t)$=$r_0(t)B_0({\bf x})$+$r_1(t)B_1({\bf x})$. 
Nevertheless, it has slightly better rms deviation from the true (spatially integrated) light curve  
(Figure \ref{sim}f top) than a corresponding average (Figure \ref{sim}f dotted line). Surprisingly, violation of the model assumption (\ref{model}) 
does not degrade the result below the quality of a simple average. The iterative convergence to the maximum-${\sf P}$
solution is demonstrated in Figure \ref{sim}f, referring to simulation \ref{sim}c). The total probability (boldface) 
is dominated by the log likelihood (solid line), indicating that the demodulation is primary determined by the
agreement with the observation.

\begin{figure}[ht] 
\centerline{\includegraphics[width=170mm]{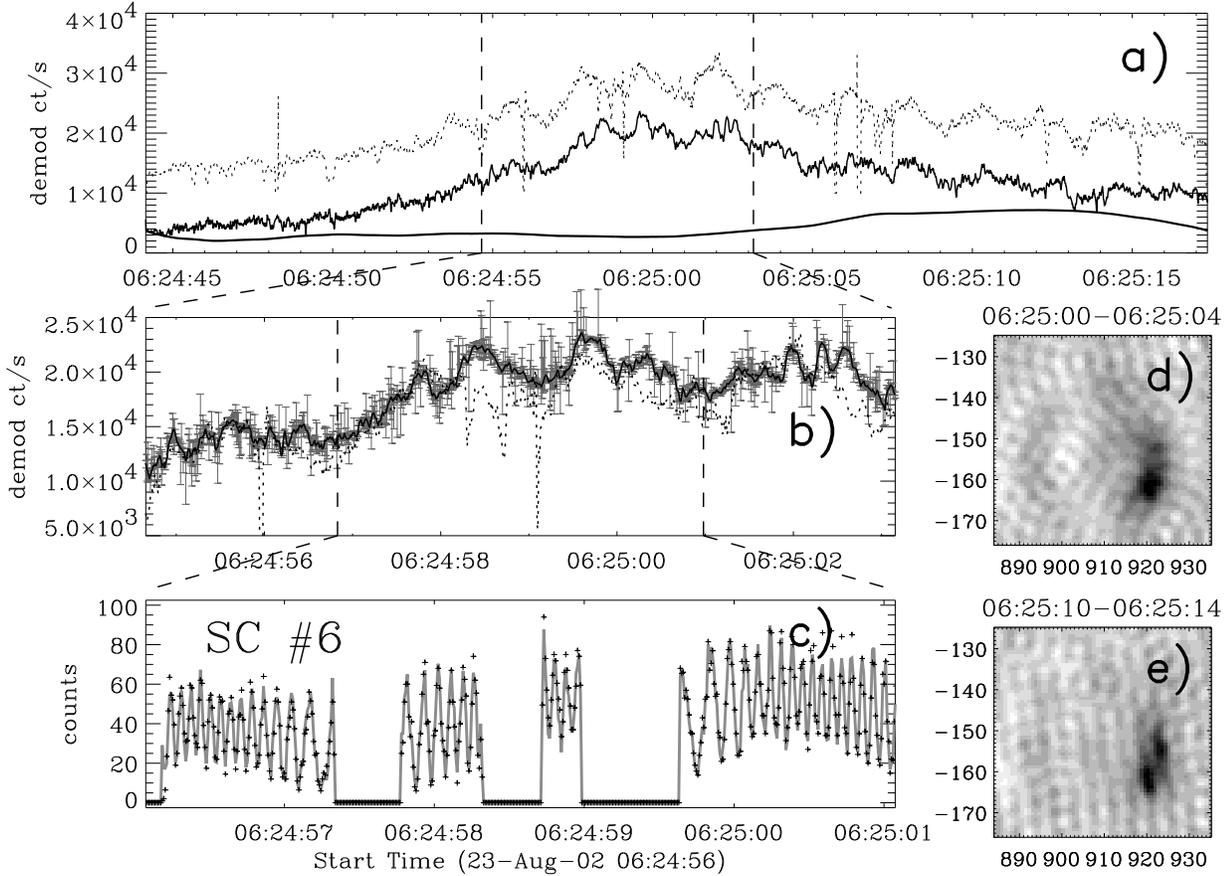}}
\vspace{-5mm}
\caption{Flare of August 23, 2002, 06:25 UT: a) average over subcollimators 1,3,4 and time $\tau_a$=0.08s 
(dotted, shifted for better clarity), demodulation $r_{0,t}$+$r_{1,t}$ (solid line), and gradual component
$r_{0,t}$ (smooth curve) from subcollimators 1,3,4,6,7,8,9; b) close-up of demodulation (solid) with
errors (gray, see text) and moving average (dotted); c) example of observed (crosses) and predicted (gray line) counts 
for subcollimator 6. e), d): CLEAN images at 
total, gradual maxima (axes in heliocentric arc seconds).} 
\label{demod_result} 
\end{figure} 
 
The algorithm is then applied to RHESSI observations of solare flares.
A flare with rich temporal fine structures occurred on August 23, 2002, 06:25 UT (Figure \ref{demod_result}). 
The energy band under consideration is 6-25 keV, and the mean observed count rate is
1700/s/subcollimator. The time bins are 9.78ms, so that there are $\ga$10 counts per time bin. The flare center 
is at (920'',-160'') (Figure \ref{demod_result}d-e). The time scale 
achievable in the demodulation is in the order of 0.1s, by the order-of-magnitude argument that for each 
(temporal+spatial) degree of freedom there should be some 100 photons.
A first estimate on the true light curve is provided by the average $\langle c \rangle/\langle a_0 L \rangle$ 
over subcollimators 1,3,4 and time $\tau_a$=0.08s (Figure \ref{demod_result}a-b, dotted, shifted for better clarity). 
The demodulation (solid line) is obtained from all subcollimators except \#2 and \#5, which are 
excluded because of high background (\#2) (Smith et al., 2003), and a distinctly 
inferior likelihood (\#5) which is not fully understood at present.
The decomposition into impulsive and gradual components (Eq. \ref{model}) removes data 
gaps, and also shows that the gradual component appears delayed (Figure \ref{demod_result}a,  
smooth curve). This seems consistent with a spatial change of the brightness distribution 
(Figure \ref{demod_result}d-e), suggesting the emergence of a gradual (possibly thermal) source 
at $\sim$(924'',-155''). Figure \ref{demod_result}c shows an example of observed counts (crosses),  
together with the predicted Poisson parameter (gray line). The data gaps have zero counts.
The correlation time of $r_{1,t}$ derived from the autocorrelation
is $\tau_1$ = 0.163s; the estimate (Eq. \ref{tau}) gives 0.164s. An important issue is 
the reliability of the demodulated fine structures. This can be tested by locally
perturbing the maximum-${\sf P}$ solution $r_{1,t}$ until $\log {\sf L}$ changes by a given
amount (Eadie et al. 1971). Adopting the conventional threshold $\Delta \log {\sf L}$ = $\frac{1}{2}$,
one obtains the error bars shown in Figure \ref{demod_result}b. The average error is 1400 ct/s. Large 
error bars are characteristic for times with data gaps in several subcollimators.
   
Let us finish this discussion with a last but important point.
Although the formalism of Eq. (\ref{F}) allows for arbitrarily small count rates, there is
no gain in temporal resolution if $\Delta t$ is made so small that only few 
(or fractions of) photons are contained in a time bin. At such low count rates, the Poisson
error is large, and Eq. (\ref{F}) becomes dominated by the smoothness constraint. Higher time resolution can only
be achieved at higher count rates, or at the expense of statistical significance ($\chi^2_{red} < 1$). 
As a conservative rule-of-thumb, a few $10^4$ counts/s are needed to resolve 100 ms structures on the 5\% level.
It is not easy to beat this limit, not even for extremely transient and bright events, because
their modulation is hard to estimate and dominates the Poisson error.

\section*{ACKNOWLEDGEMENTS} 
 
The author thanks G. Hurford and R. Schwartz for helpful discussions.

\vspace{3mm} 
 
E-mail address: ~~~ arzner@astro.phys.ethz.ch

Manuscript received 5 December 2002; revised 20 January 2004; accepted 22 January 2004
 
\end{document}